\documentclass[prd, amsmath, amssymb, showpacs, superscriptaddress,
nofootinbib, twocolumn]{revtex4-1}  
\usepackage{graphicx}
\usepackage{dcolumn}
\usepackage{bm}
\usepackage{rotating}
\usepackage{multirow}
\usepackage{overpic}
\usepackage{hhline}
\usepackage{amssymb}   
\usepackage[colorlinks,
linkcolor=blue,
anchorcolor=black,
citecolor=blue
]{hyperref}
\normalsize
\begin{document}
\title{\boldmath Measurements of the branching fractions of
\texorpdfstring{$\eta_c\to$}{etac->}
\texorpdfstring{$K^+K^-\pi^0$}{KKpi0},
\texorpdfstring{$K^0_S K^{\pm}\pi^{\mp}$}{KsKpi},
\texorpdfstring{$2(\pi^+\pi^-\pi^0)$}{6pi},
and \texorpdfstring{$p \bar{p}$}{ppbar}}
\author{
\begin{small}
\begin{center}
M.~Ablikim$^{1}$, M.~N.~Achasov$^{10,d}$, S. ~Ahmed$^{15}$, M.~Albrecht$^{4}$, M.~Alekseev$^{55A,55C}$, A.~Amoroso$^{55A,55C}$, F.~F.~An$^{1}$, Q.~An$^{52,42}$, Y.~Bai$^{41}$, O.~Bakina$^{27}$
% ...
\\
\vspace{0.2cm}
(BESIII Collaboration)\\
\vspace{0.2cm} {\it
$^{1}$ Institute of High Energy Physics, Beijing 100049, People's Republic of China\\
% ...
$^{4}$ Bochum Ruhr-University, D-44780 Bochum, Germany\\
% ...
$^{10}$ G.I. Budker Institute of Nuclear Physics SB RAS (BINP), Novosibirsk 630090, Russia\\
% ...
$^{15}$ Helmholtz Institute Mainz, Johann-Joachim-Becher-Weg 45, D-55099 Mainz, Germany\\
% ...
$^{27}$ Joint Institute for Nuclear Research, 141980 Dubna, Moscow region, Russia\\
% ...
$^{41}$ Southeast University, Nanjing 211100, People's Republic of China\\
$^{42}$ State Key Laboratory of Particle Detection and Electronics, Beijing 100049, Hefei 230026, People's Republic of China\\
% ...
$^{52}$ University of Science and Technology of China, Hefei 230026, People's Republic of China\\
% ...
$^{55}$ (A)University of Turin, I-10125, Turin, Italy; (B)University of Eastern Piedmont, I-15121, Alessandria, Italy; (C)INFN, I-10125, Turin, Italy\\
% ...
$^{d}$ Also at the Novosibirsk State University, Novosibirsk, 630090, Russia\\
% ...
}\end{center}
\end{small}
}
\noaffiliation{}
\date{\today}
\begin{abstract}
  Using data samples collected with the BESIII detector
  at center-of-mass energies $\sqrt{s} = 4.23, 4.26, 4.36,$ and $4.42$~\rm{GeV},
  we measure the branching fractions of
  $\eta_c\to K^+K^-\pi^0$, $K^0_S K^{\pm}\pi^{\mp}$,
  $2(\pi^+\pi^-\pi^0)$, and $p \bar{p}$,
  via the process $e^+e^-\to\pi^+\pi^-h_c$, $h_c\to\gamma\eta_c$.
  The corresponding results are
  $(1.15\pm0.12\pm0.10)\%$, $(2.60\pm0.21\pm0.20)\%$,
  $(15.2\pm1.8\pm1.7)\%$, and $(0.120\pm0.026\pm0.015)\%$, respectively.
  Here the first uncertainties are statistical, and the second ones systematic.
  Additionally, the charged track multiplicity of $\eta_c$ decays
  is measured for the first time.
\end{abstract}
\pacs{12.38.Qk, 14.40.Pq, 13.25.Gv}
\maketitle
\section{ \boldmath INTRODUCTION}
\label{sec:introduction}
Many new charmonium or charmonium-like states have been discovered recently~\cite{xyz_states},
which broaden our horizon on understanding the charmonium family.
These states have led to a revived interest
in improving the quark-model picture of hadrons.
However, the knowledge of the lowest lying charmonium state, $\eta_c$,
is relatively poor compared to the other charmonium states.
The reason is that most of the measurements involving $\eta_c$ were performed
using {\it M}1 transitions from $J/\psi$ or hindered {\it M}1
transitions from $\psi(3686)$.
In these decays, the interference between $\eta_c$ and non-$\eta_c$ amplitudes
affects the $\eta_c$ lineshape~\cite{getac}.
The branching fraction~(BF) of $\eta_c$ decays and the {\it M}1 transition rate
are entangled.
The insufficient understanding of the $\eta_c$ properties has so far
prevented precise studies of $\eta_c$ decays themselves or of decays involving the $\eta_c$.
For example, in 2002, the Belle Collaboration release the measurements
on the total cross section of the exclusive production of $J/\psi+\eta_c$
via the $e^+e^-$ annihilation at the center-of-mass collision energy
$\sqrt{s}=10.58~{\rm GeV}$ ~\cite{multiplicity}
with the result of
$\sigma[e^+e^-\to J/\psi+\eta_c]\times {\rm BF}(\eta_c\to \ge 4~{\rm charged})=33^{+7}_{-6}\pm 9~{\rm fb}$.
These measurements were improved as 
$\sigma[e^+e^-\to J/\psi\eta_c(\gamma)]\times
{\rm BF}(\eta_c\to\ge 2~\rm charged)=25.6\pm 2.8\pm 3.4~fb$ ~\cite{multiplicity_2}.
In 2005, the BABAR Collaboration independently measured the total cross section
as $17.6\pm 2.8^{+1.5}_{-2.1}~{\rm fb}$~\cite{multiplicity_3}.
As the number of charged tracks is required in these measurements,
the results will be improved if the charged tracks multiplicity is fully studied.

Recently, the {\it E}1 transition $h_c\to\gamma\eta_c$ was found
to be a perfect process to measure
both $\eta_c$ resonant parameters and its decay BFs~\cite{guo_aiqiang_etac}.
In addition, the $h_c$ production proceeds via $\psi(3686)\to\pi^0 h_c$,
where the interference effect between $\eta_c$ and non-$\eta_c$
is much less than that in $J/\psi,\psi(3686)$ radiative transition.
One can draw such a conclusion according to the following calculation.
The $E$1 transition rate, ${\rm BF}(h_c\to\gamma \eta_c)=50\%$,
is about 2 orders of magnitude larger
than that of the $M$1 transition ${\rm BF}(\psi(3686)\to\gamma \eta_c) =  0.3\%$~\cite{pdg}.
On the other hand, the background that can interfere with the signal
comes from charmonium radiative decays,
\emph{e.g.} $h_c, \psi(3686)\to\gamma + {\rm hadrons}$.
If we assume the radiative decay rates of $h_c$ and $\psi(3686)$
to be at the same level,
therefore, this kind of background in the process  $h_c\to\gamma \eta_c$
should be 1 to 2 orders of magnitude less than 
in  $\psi(3686)\to\gamma \eta_c$.

BESIII has collected sizable data samples between 4.009 and 4.600 GeV
(called ``XYZ data'' hereafter) since 2013
to study the XYZ states~\cite{xyz_luminosities}.
A large production rate of $e^+e^-\to\pi^+\pi^-h_c$ has been found~\cite{guo_yuping_zc}.
The total number of $h_c$ events in all these data samples combined is comparable to
that from $\psi(3686)\to\pi^0h_c$ decays in BESIII data, according to the measured cross section
and the corresponding integrated luminosity at each energy point.
The $h_c$ is tagged by the recoil mass ($RM$) of $\pi^+\pi^-$ in XYZ data,
while it is tagged by the recoil mass of $\pi^0$ in $\psi(3686)$ data. 
Generally, the two-charged-pion mode
has lower background and higher detection efficiency than the neutral pion mode.

In this paper, we report a measurement of the BFs
of four $\eta_c$ exclusive decays
via the process $e^+e^-\to \pi^+\pi^-h_c$, $h_c\to\gamma\eta_c$.
These exclusive decays are
$\eta_c\to K^{+}K^{-}\pi^0$, $K^0_SK^{\pm}\pi^{\mp}$,
$2(\pi^+\pi^-\pi^0)$, and $p\bar{p}$, respectively.

Apart from the BF measurement mentioned above,
we also measure the charged tracks multiplicities
in inclusive $\eta_c$ decays by using an unfolding method~\cite{N_psip}.

\section{METHODOLOGY}
The BFs of $\eta_c$ exclusive decays are obtained
by a simultaneous fit to the $RM$ spectrum of $\pi^+\pi^-\gamma$
for both inclusive and exclusive modes.
The BFs are common parameters independent of the center of mass energy.
The numbers of the $\eta_c$ signal events
of the exclusive and inclusive decay modes
can be calculated by the following formulas,
\begin{widetext}
  \begin{equation}
    N^i_{\rm exclusive} = 
      \mathcal{L}^{i}\times\sigma^{i}(e^+e^-\to\pi^+\pi^- h_c)
      \times{\rm BF}(h_c\to \gamma\eta_c)\times {\rm BF}(\eta_c\to X)
      \times {\rm BF}(X\to Y)\times\epsilon^{i}_{\rm exclusive},
    \label{Eq:event_number_exclusive}
  \end{equation}
    and
  \begin{equation}
      N^i_{\rm inclusive} = 
      \mathcal{L}^{i}\times\sigma^{i}(e^+e^-\to\pi^+\pi^- h_c)
      \times{\rm BF}(h_c\to \gamma\eta_c)
      \times\epsilon^{i}_{\rm inclusive},
    \label{Eq:event_number_inclusive}
  \end{equation}
\end{widetext}
where
the subscript $i$ denotes the different center-of-mass energy points.
$\mathcal{L}$ and $\sigma$ denote the luminosity and cross section, respectively.
$X$ denotes a certain $\eta_c$ exclusive decay mode,
$Y$ denotes the possible $\pi^0$ or $K^0_S$ final state from $X$ decay.
$\epsilon$ denotes the detection efficiency
determined by Monte Carlo~(MC) simulations.

By comparing Eq.~\eqref{Eq:event_number_exclusive} and Eq.~\eqref{Eq:event_number_inclusive},
${\rm BF}(\eta_c\to X)$ can be extracted as
\begin{widetext}
  \begin{equation}
      {\rm BF}(\eta_c\to X) =
      \frac{N^{i}_{\rm exclusive}/
      \Big({\rm BF}(X\to Y)\times\epsilon^{i}_{\rm exclusive}\Big)}
      {N^{i}_{\rm inclusive}/\epsilon^{i}_{\rm inclusive}}.
      \label{Eq:event_number}
  \end{equation}
\end{widetext}

In the simultaneous fit, the total number of free parameters is less than
in the fits taken individually, due to common parameters such as the $\eta_c$ mass and width,
\emph{etc.}
In addition, some parameters,
for example, $\sigma(e^+e^-\to\pi^+\pi^-h_c)$, ${\cal L}$,
are not necessary in the measurement according to Eq.~\eqref{Eq:event_number}, resulting in reduced
statistical uncertainties.
In addition, systematic uncertainties from the same sources,
\emph{e.g.}, the tracking efficiency of two pions from $e^+e^-\to\pi^+\pi^-h_c$,
can be canceled.
\section{\boldmath DETECTOR AND DATA SAMPLES}
\label{sec:detector_and_data_samples}

The BESIII detector is a magnetic spectrometer~\cite{Ablikim:2009aa}
located at the Beijing Electron Positron Collider~(BEPCII)~\cite{Yu:IPAC2016-TUYA01}.
The cylindrical core of the BESIII detector consists of a helium-based
multilayer drift chamber (MDC),
a plastic scintillator time-of-flight system (TOF),
and a CsI(Tl) electromagnetic calorimeter (EMC),
which are all enclosed in a superconducting solenoidal magnet
providing a 1.0~T magnetic field. The solenoid is supported by an
octagonal flux-return yoke with resistive plate counter muon
identifier modules interleaved with steel. The acceptance of
charged particles and photons is 93\% over $4\pi$ solid angle. The
charged-particle momentum resolution at $1~{\rm GeV}/c$ is
$0.5\%$, and the specific energy loss (${\rm d}E/{\rm d}x$) resolution is $6\%$ for the electrons
from Bhabha scattering. The EMC measures photon energies with a
resolution of $2.5\%$ ($5\%$) at $1$~GeV in the barrel (end cap)
region. The time resolution of the TOF barrel part is 68~ps, while
that of the end cap part is 110~ps.

The data samples collected at 4 center-of-mass energies, i.e.
$\sqrt{s} = 4.23$, $4.26$, $4.36$, and $4.42$~\rm{GeV}~\cite{xyz_luminosities},
are used for our studies.
Simulated samples produced with the {\sc
geant4}-based~\cite{geant4} MC package which
includes the geometric description of the BESIII detector and the
detector response, are used to determine the detection efficiency
and to estimate the backgrounds. The simulation includes the beam
energy spread and initial state radiation (ISR) in the $e^+e^-$
annihilations modeled with the generator {\sc kkmc}~\cite{ref:kkmc}.

The inclusive MC samples with equivalent luminosities the same as the data samples
consist of the production of open charm
processes, the ISR production of vector charmonium(-like) states,
and the continuum processes incorporated in {\sc
kkmc}~\cite{ref:kkmc}.
The known decay modes are modeled with {\sc evtgen}~\cite{ref:evtgen}
using branching fractions taken from PDG~\cite{pdg},
and the remaining unknown decays
from the charmonium states with {\sc lundcharm}~\cite{ref:lundcharm}.
The final state radiations (FSR)
from charged final state particles are incorporated with the {\sc
photos} package~\cite{photos}.

Signal MC samples with 200\,000 events each are generated for each $\eta_c$ decay mode
(inclusive and exclusive decays)
at each center-of-mass energy.
ISR is simulated using {\sc kkmc} with a maximum energy
for the ISR photon corresponding to the $\pi^+\pi^-h_c$ mass threshold.
The {\it E}1 transition $h_c\to\gamma\eta_c$ is generated
with an angular distribution of $1+\cos^2\theta$,
where $\theta$ is the angle of the {\it E}1 photon
with respect to the $h_c$ helicity direction in the $h_c$ rest frame.
The inclusive decays of $\eta_c$ are produced
similarly to the inclusive MC samples.

\section{\boldmath EVENT SELECTIONS}
In this analysis, the $\eta_c$ signal is tagged with $RM(\pi^+\pi^-\gamma)$
by requiring $RM(\pi^+\pi^-)$ in $h_c$ signal region.
For the inclusive mode, at least two charged tracks and one photon is required.
For the exclusive modes, the requirements on charged tracks and photon candidates
depend on their respective final state.

Charged tracks at BESIII are reconstructed from MDC hits
within a polar-angle~($\theta$) acceptance range of $|\cos\theta|<0.93$.
We require that these tracks pass
within 10~cm of the interaction point in the beam direction
and within 1~cm in the plane perpendicular to the beam.
Tracks used in reconstructing $K^0_S$ decays are exempted from these requirements.

A vertex fit constrains charged tracks to a common production vertex,
which is updated on a run-by-run basis.
For each charged track, TOF and ${\rm d}E/{\rm d}x$ information
is combined to compute particle identification~(PID) confidence levels
for the pion, kaon, and proton hypotheses.

Electromagnetic showers are reconstructed by clustering EMC crystal energies.
Efficiency and energy resolution are improved
by including energy deposits in nearby TOF counters.
A photon candidate is defined as an isolated shower with an energy deposit
of at least 25~MeV in the barrel region~($|\cos\theta|<0.8$),
or of at least 50~MeV in the end-cap region~($0.86<|\cos\theta|<0.92$).
Showers in the transition region between the barrel and the end-cap
are not well measured and are rejected.
An additional requirement on the EMC hit timing suppresses electronic noise
and energy deposits unrelated to the event.

A candidate $\pi^0$ is reconstructed from pairs of photons
with an invariant mass in the range $|M_{\gamma\gamma}-m_{\pi^0}|<15$ MeV/c$^2$~\cite{pdg}.
A one-constraint~(1-C) kinematic fit is performed to improve the energy resolution,
with the $M_{\gamma\gamma}$ constrained to the known $\pi^0$ mass.

We reconstruct $K^0_S\to\pi^+\pi^-$ candidates
using pairs of oppositely charged tracks with an invariant mass in the range
$|M_{\pi^+\pi^-}-m_{K^0_S}|<20$~MeV/c$^2$,
where $m_{K^0_S}$ is the known $K^0_S$ mass~\cite{pdg}.
To reject random $\pi^+\pi^-$ combinations,
a secondary-vertex fitting algorithm is employed
to impose the kinematic constraint between the production and decay vertices~\cite{vertex}.
Accepted $K^0_S$ candidates are required to have a decay length of at least twice the vertex resolution.
If there is more than one $\pi^+\pi^-$ combinations in an events,
the one with the smallest $\chi^2$ of the secondary vertex fit is retained.

\begin{figure*}[!htp]
  \begin{center}
    \begin{overpic}[width=0.40\textwidth]
      {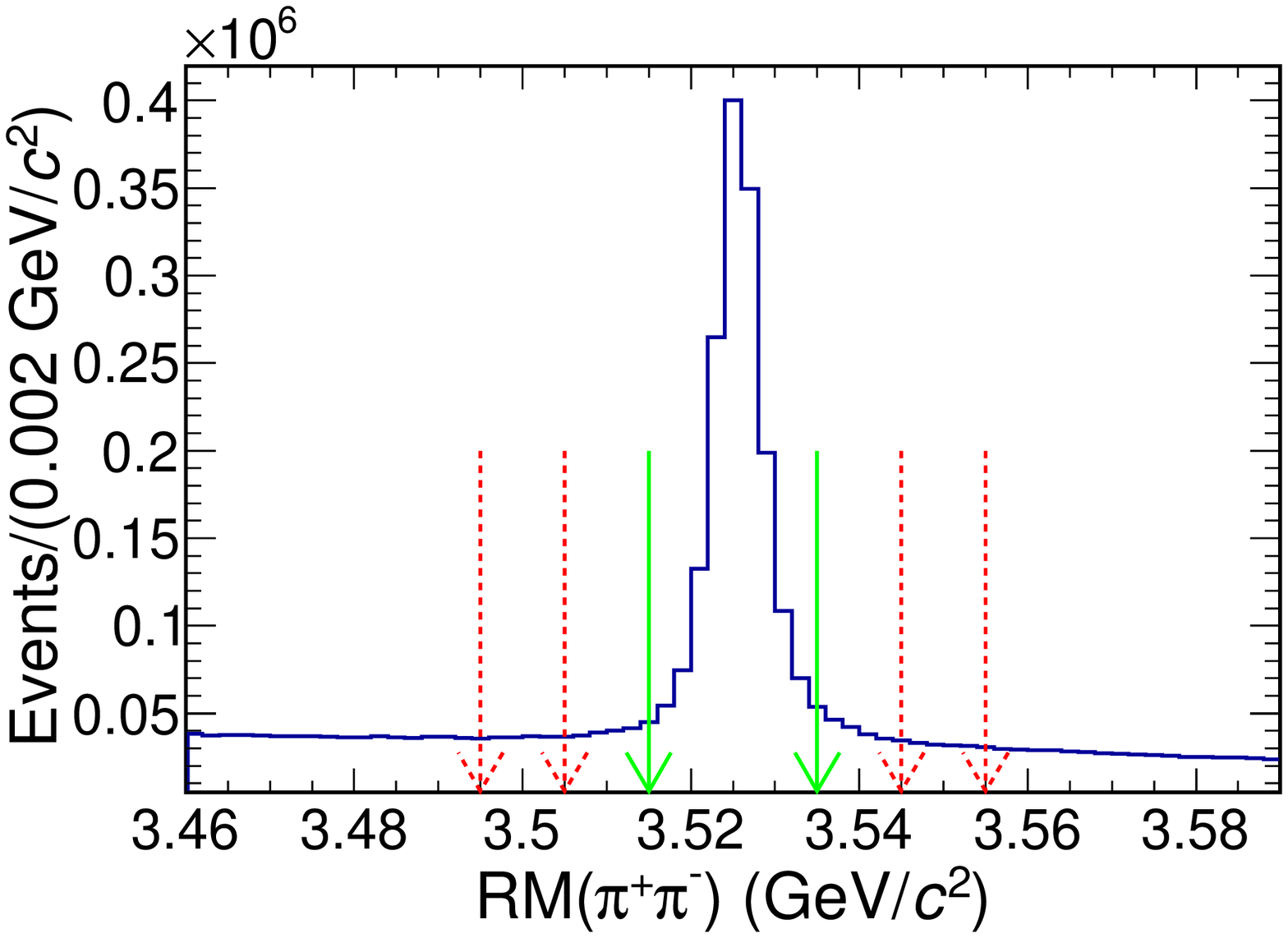}
      \put(80,55){(a)}
    \end{overpic}
    \begin{overpic}[width=0.40\textwidth]
      {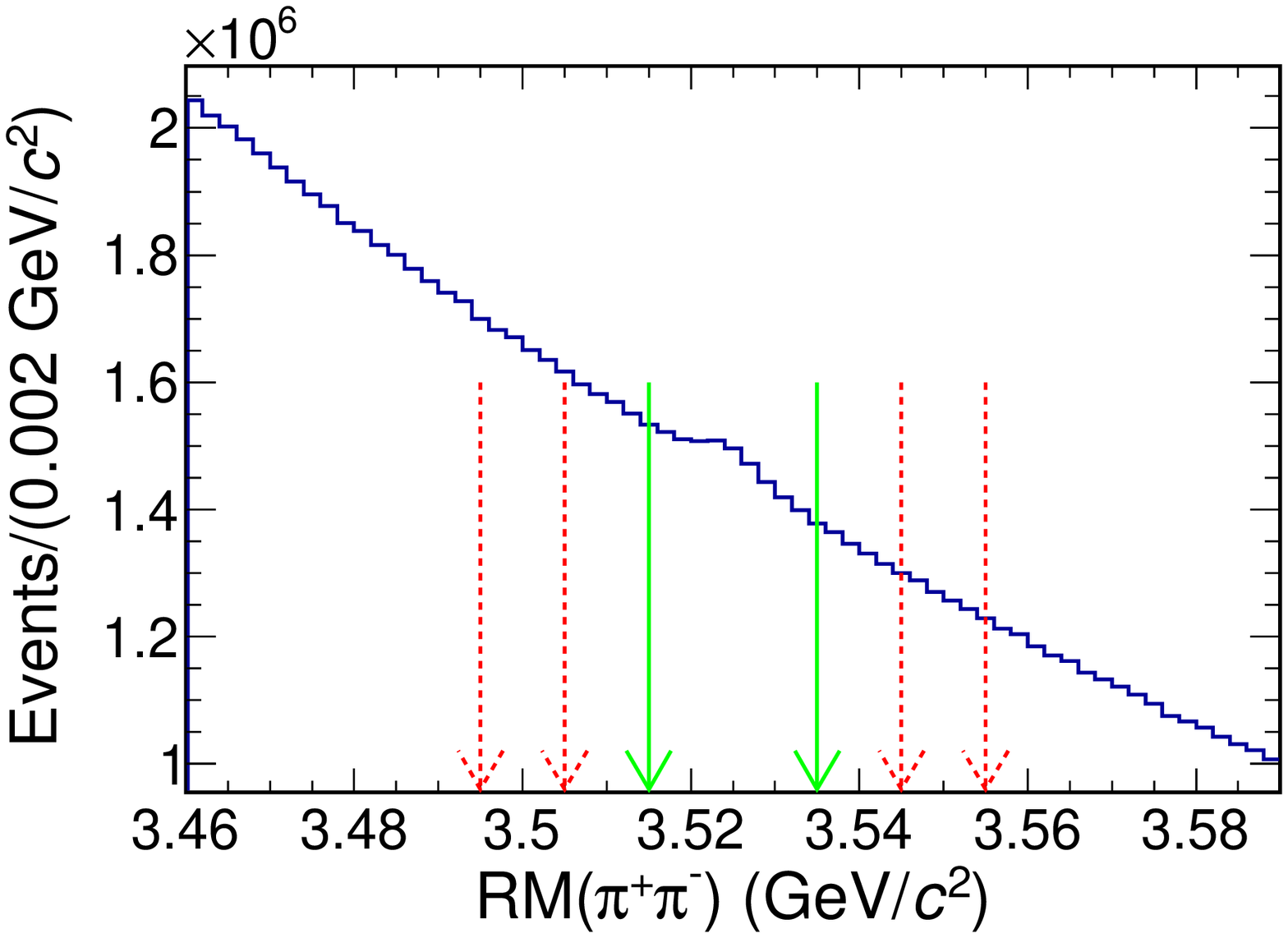}
      \put(80,55){(b)}
    \end{overpic}
  \end{center}
  \caption{Distribution of $RM(\pi^+\pi^-)$
  of the $\eta_c$ inclusive decay from
  signal MC simulation~(a) and data~(b)
  summed over all the four center-of-mass energies.
  The $h_c$ signal and sideband regions are marked
  by the solid and dashed arrows, respectively. }
  \label{FIG:recoil_pipi_inc}
\end{figure*}

In selecting the candidates of the $\eta_c$ inclusive decay,
all charged tracks are assumed to be pions,
and events with at least one combination
satisfying $RM(\pi^+\pi^-)\in[3.46, 3.59]~{\rm GeV}/c^2$
and $RM(\pi^+\pi^-\gamma)\in[2.52, 3.4]~{\rm GeV}/c^2$
are kept for further analysis.
The region satisfying $RM(\pi^+\pi^-)\in[3.515,~3.535]$~GeV/$c^2$
is taken as the $h_c$ signal region,
while the regions satisfying
$RM(\pi^+\pi^-)\in [3.495,~3.505]$~GeV/$c^2$ or
$RM(\pi^+\pi^-)\in [3.545,~3.555]$~GeV/$c^2$
are taken as the $h_c$ sidebands region.
Figure~\ref{FIG:recoil_pipi_inc} shows
the distribution of $RM(\pi^+\pi^-)$ for all $\pi^+\pi^-$ combinations
from the inclusive decay mode in signal MC and data (summed over four center-of-mass energies), respectively.

\begin{table}[!htbp]
  \begin{center}
    \caption{\small Requirements of the number of photons, charged tracks, $\pi^0$,
    and $K^0_S$ candidates in exclusive $\eta_c$ decay modes,
    denoted as $N_{\rm charge}$, $N_{\gamma}$, $N_{\pi^0}$, and $N_{K^0_S}$,
    respectively.}
    \label{table:exclusive_requirements}
    \begin{tabular}{c|c|c|c}
      \hline
      \hline
      Decay mode & $N_{\rm charge}$ & $N_{\gamma}$ & Other requirements \\
      \hline
      $\eta_c\to K^+K^-\pi^0$ & $=2$ & $\ge 3$ & $N_{\pi^0}\ge 1$ \\
      $\eta_c \to K_S^0 K^{\pm}\pi^{\mp}$ & $=4$ & $\ge 1$ & $N_{K^0_S}=1$ \\
      $\eta_c \to 2(\pi^+\pi^-\pi^0)$ & $=4$ & $\ge 5$ & $N_{\pi^0}\ge 2$ \\
      $\eta_c \to p\bar{p}$ & $=2$ & $\ge 1$ & - \\
      \hline
      \hline
    \end{tabular}
  \end{center}
\end{table}
For the selection of exclusive $\eta_c$ decays,
the requirements on the number of photons and charged tracks are listed
in Table~\ref{table:exclusive_requirements}.
A four-constraint (4C) kinematic fit imposing overall energy-momentum conservation is performed.
To determine the species of final state particles
and to select the best combination
when additional photons~(or $\pi^0$ candidates) are found in an event,
the combination with the minimum value of
$\chi^2=\chi^2_{\rm{4C}} + \chi^2_{\rm{1C}} +
\sum^{N_{\rm charge}}_{i=1}\chi^2_{\rm{PID}} + \chi^2_{\rm{Vertex}}$
is selected for further analysis, where
$\chi^2_{\rm{4C}}$ is the $\chi^2$
from the four-momentum conservation kinematic fit
and $\chi^2_{\rm{1C}}$ is the sum
of the 1C~(mass constraint of the two daughter photons) $\chi^2$
of the $\pi^0$ in the final state.
$\chi^2_{\rm{PID}}$ is the $\chi^2$ from PID of different particle hypothesis,
using the energy loss in the MDC and the time measured with the TOF system,
$N_{\rm charge}$ is the number of the charged tracks in the final states.
$\chi^2_{\rm Vertex}$ is the $\chi^2$ of the vertex fit
in $K^0_S$ reconstruction.
The $\chi^2_{\rm 4C}$ is required to be not more than 50
depending on the $\eta_c$ decay modes,
which is optimized using the figure of merit
$N_S/\sqrt{N_S + N_B}$,
where $N_S$ is the number of signal events obtained from
MC simulation (normalized to data luminosity),
while $N_B$ is the number of background events obtained from
the sidebands of $h_c$ in data.
\begin{table}[!htbp]\small
  \begin{center}
    \caption{The requirements of $\chi^2_{\rm 4C}$ for the exclusive decays of
    $\eta_c$.}
    \label{table:chisq_4c_cut}
    \begin{tabular}{c|c|c|c|c}
      \hline
      \hline
      $\sqrt{s}$ (GeV)& $K^0_S K^{\pm}\pi^{\mp} $& $K^+ K^- \pi^0 $ & $2(\pi^+\pi^-\pi^0)$ & $p \bar{p} $ \\
      \hline
      4.23 & 45 & 25 & 35 & 40 \\
      \hline
      4.26 & 45 & 15 & 30 & 40 \\
      \hline
      4.36 & 45 & 25 & 25 & 40 \\
      \hline
      4.42 & 50 & 20 & 35 & 40 \\
      \hline
      \hline
    \end{tabular}
  \end{center}
\end{table}
The requirement on $\chi^2_{\rm 4C}$ for the different exclusive decay modes are
listed in Table~\ref{table:chisq_4c_cut}.
In addition, we require the same $h_c$ mass windows
on the $RM(\pi^+\pi^-)$ spectra for both inclusive and exclusive modes.

\section{\boldmath Numerical results of ${\rm BF}(\eta_c\to X)$}
\begin{figure*}[!htpb]
  \begin{center}
    \begin{overpic}[width=0.40\textwidth]
      {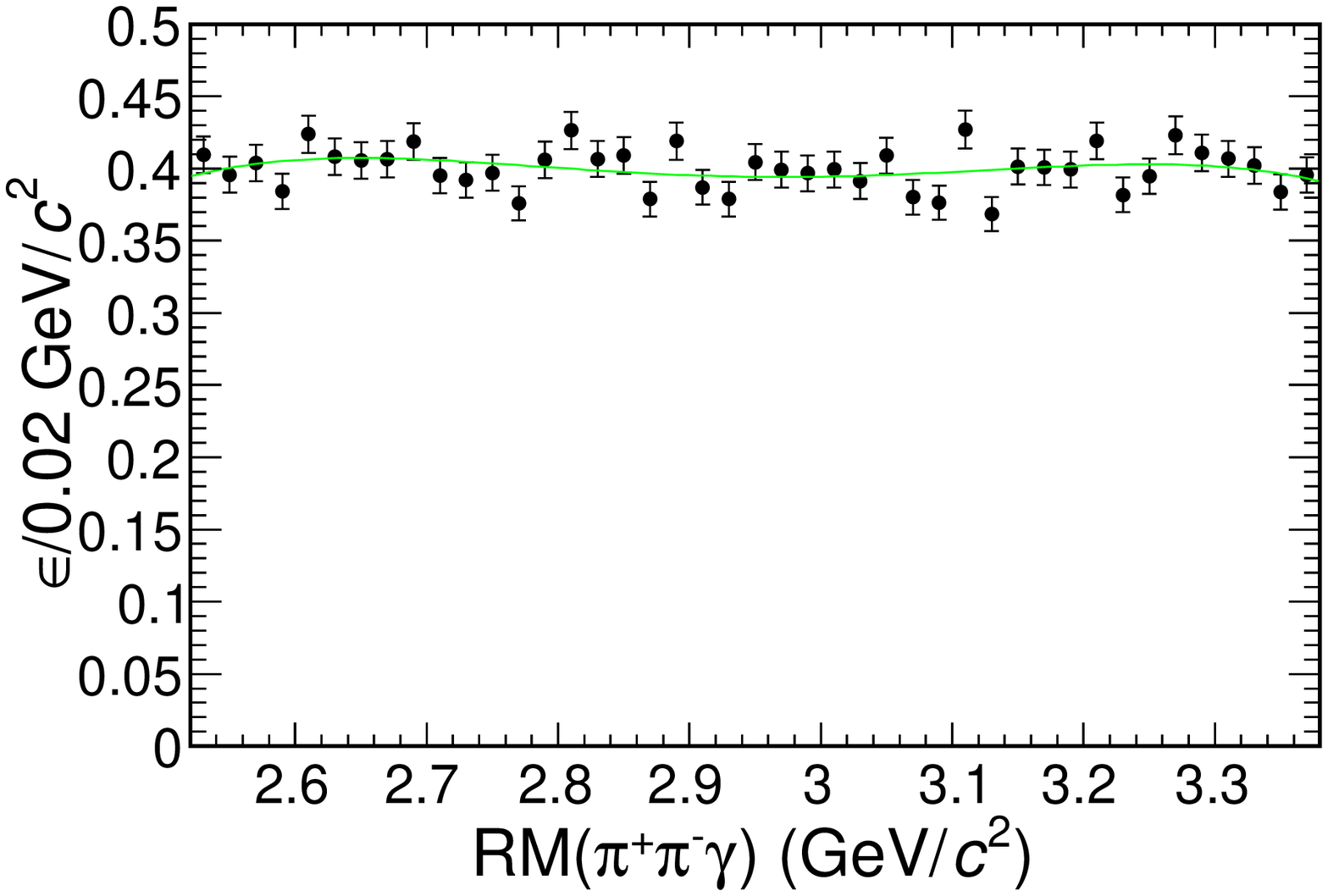}
      \put(75,40) {(a)}
    \end{overpic}
    \begin{overpic}[width=0.40\textwidth]
      {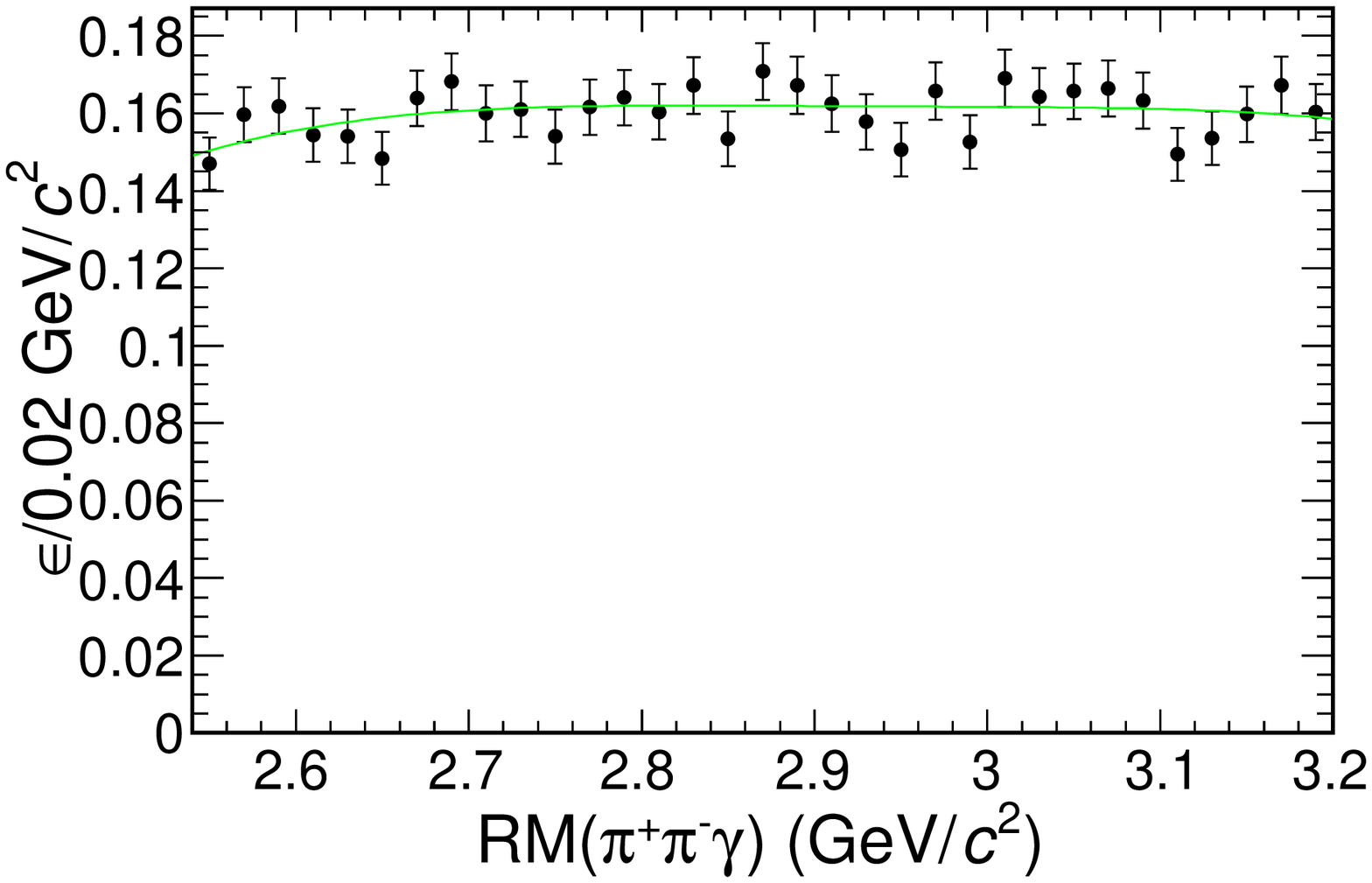}
      \put(75,40){(b)}
    \end{overpic}
  \end{center}
  \caption{Efficiencies along the $RM(\pi^+\pi^-\gamma)$ spectra
  from MC simulation at $\sqrt{s}=$4.23~GeV for
  inclusive decay~(a) and $\eta_c\to K^+K^-\pi^0$~(b).
  The curves are the fit results.}
  \label{FIG:efficiency_curves}
\end{figure*}

A simultaneous unbinned maximum likelihood fit to the $\mathrm{RM}(\pi^+\pi^-\gamma)$ spectrum
of the exclusive decays and the inclusive decay of $\eta_c$ at the four center-of-mass energies
is performed to obtain the branching fractions ${\rm BF}(\eta_c\to X)$.
The fit function is parameterized as follows:
\begin{widetext}
  \begin{equation}
    F(M) = \sigma\otimes[\epsilon(M)\times|BW(M)|^2\times E^3_{\gamma}\times f_{\rm d}(E_{\gamma})] + B(M),
  \end{equation}
\end{widetext}
where the signal function is described by a Breit-Wigner function, $BW(M)$,
convolved with the detection resolution, $\sigma$.
The mass and width of $BW(M)$ are fixed to the $\eta_c$ nominal values taken from the PDG~\cite{pdg}.
$M$ represents the recoil mass $RM(\pi^+\pi^-\gamma)$.
The detection resolution is described by a double Gaussian function,
whose parameters are obtained from MC simulations.
$\epsilon(M)$ is the efficiency curve, obtained from
a fit of the efficiencies along the $RM(\pi^+\pi^-\gamma)$ spectrum with
a polynomial function and fixed in the fit to data.
Figure~\ref{FIG:efficiency_curves} shows the efficiencies
along the $RM(\pi^+\pi^-\gamma)$ spectrum
for the inclusive $\eta_c$ decay and the exclusive decay $\eta_c\to K^+K^-\pi^0$
at $\sqrt{s} = 4.23~{\rm GeV}$.

$E_\gamma = (m_{h_{c}}^2 - M^2)/2m_{h_{c}}$
is the energy of the transition photon,
where $m_{h_c}$ is the $h_c$ mass~\cite{pdg}.
\begin{center}
  $f_{\rm d}(E_{\gamma}) = \frac{E^2_0}{E_{\gamma}E_0 + ( E_{\gamma}-E_0 )^2}$
\end{center}
is the damping factor~\cite{damping_factor_KEDR},
where $E_0=E_\gamma(m_{\eta_c})$ is the most probable transition energy.

$B(M)$ denotes the function which is used to describe the background shape.
For an exclusive decay mode, a polynomial function is used.
For the inclusive decay mode,
it is a combination of the distribution from $h_c$ sidebands and a polynomial function.

\begin{figure*}[!htpb]
  \begin{center}
    \begin{overpic}[width=0.95\textwidth]
      {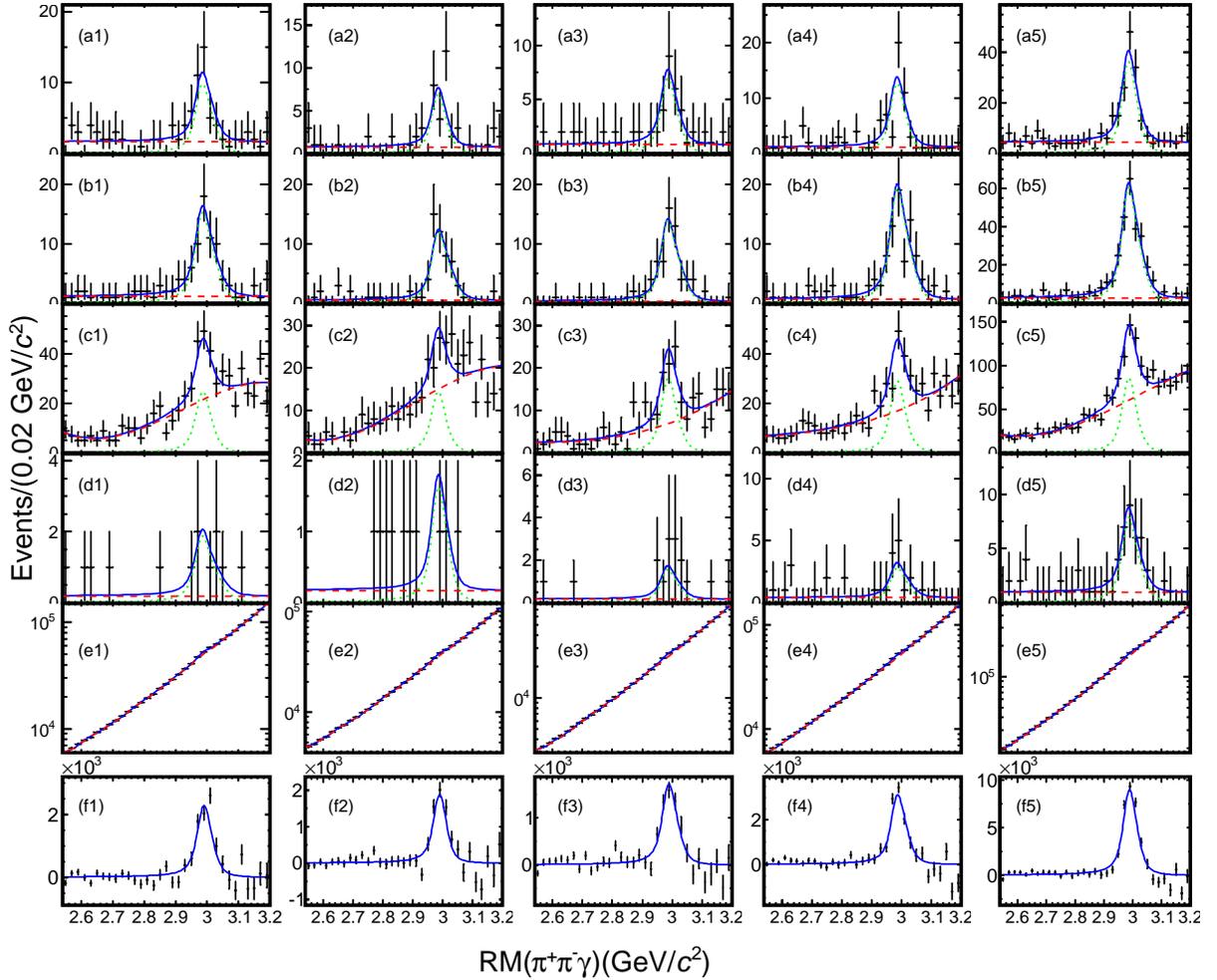}
    \end{overpic}
  \end{center}
  \caption{Projections of the simultaneous fit to data.
  The dots with error bars denote data,
  the dashed lines denote backgrounds,
  the dotted lines denote signals,
  and the solid lines are the fitting curve.
  The columns from left to right, labeled from (1) to (5),
  denote $\sqrt{s} =$ 4.23, 4.26 , 4.36 , 4.42~GeV, and the sum,
  while the rows from (a) to (d) show the 4 exclusive decay modes of $\eta_c$,
  namely, $\eta_c\to K^+K^-\pi^0$,  $K^0_S K^{\pm}\pi^{\mp}$,
  $2(\pi^+\pi^-\pi^0)$,  and $p \bar{p}$.
  Row (e) shows the fit to the inclusive $\eta_c$ decay,
  while (f) denotes the background-subtracted
  $RM(\pi^+\pi^-\gamma)$ spectrum with the signal shape overlaid.}
  \label{FIG:simultaneous_fit}
\end{figure*}
Figure~\ref{FIG:simultaneous_fit} shows the simultaneous fit results.
\begin{table*}[htbp]
  \begin{center}
    \caption{\small Detection efficiencies~($\epsilon$) for $\eta_c$ inclusive and exclusive decays,
    fit results including the observed number of signal events~($N_{\rm obs}$), and the fitted BFs
    for the four $\eta_c$ exclusive decay modes. The statistical uncertainties of the observed numbers of the signal yields
    for the inclusive decay are obtained directly from the fit, while the numbers of signal events
    for the exclusive decays are calculated via Eq.~\eqref{Eq:event_number}
    rather than being obtained directly from the fit,
    so no uncertainties are provided.}
    \label{table:simultaneous_fit}
    \begin{tabular}{c|c|c|c|c}
      \hline
      \hline
      \multicolumn{2}{c|}{Category} &
      \multirow{2}{*}{$\epsilon$~(\%)} &
      \multirow{2}{*}{$N_{\rm obs}$} &
      \multirow{2}{*}{BF~(\%)} \\
      \hhline{--~~~}
      Decay modes & $\sqrt{s}~({\rm GeV})$ &  &  &  \\
      \hhline{-----}
      \multirow{5}{*}{$\eta_c\to K^+K^-\pi^0$} &
      4.23 & 15.95 & $38.6$ &
      \multirow{5}{*}{ $1.15\pm0.12$ } \\
      \hhline{~---~} & 4.26 & 15.33 & $26.6$ & \\
      \hhline{~---~} & 4.36 & 18.82 & $30.6$ & \\
      \hhline{~---~} & 4.42 & 17.92 & $50.2$ & \\
      \hhline{~---~} & sum & - & $146.0$ & \\
      \hline
      \multirow{5}{*}{$\eta_c\to K_S^0 K^{\pm}\pi^{\mp}$} &
      4.23 & 17.50 & $66.7$ &
      \multirow{5}{*}{ $2.60\pm0.21$ } \\
      \hhline{~---~} & 4.26 & 19.67 & $53.7$ \\
      \hhline{~---~} & 4.36 & 20.67 & $52.8$ \\
      \hhline{~---~} & 4.42 & 21.22 & $93.5$ \\
      \hhline{~---~} & sum & - & $266.7$ \\
      \hline
      \multirow{5}{*}{$\eta_c\to 2(\pi^+\pi^-\pi^0)$} &
      4.23 & 2.93 & $91.9$ &
      \multirow{5}{*}{ $15.2\pm1.8$ } \\
      \hhline{~---~} & 4.26 & 2.60 & $58.6$ & \\
      \hhline{~---~} & 4.36 & 3.38 & $71.2$ & \\
      \hhline{~---~} & 4.42 & 3.07 & $111.6$ & \\
      \hhline{~---~} & sum & - & $333.3$ & \\
      \hline
      \multirow{5}{*}{$\eta_c\to p\bar{p}$} &
      4.23 & 34.68 & $8.4$ &
      \multirow{5}{*}{ $0.120\pm 0.026$ } \\
      \hhline{~---~} & 4.26 & 37.67 & $7.0$ & \\
      \hhline{~---~} & 4.36 & 40.00 & $6.9$ & \\
      \hhline{~---~} & 4.42 & 40.72 & $12.1$ & \\
      \hhline{~---~} & sum & - & $34.4$ & \\
      \hline
      \multirow{4}{*}{Inclusive decays} &
      4.23 & 40.45 & 8\,314$\pm$584 &
      \multirow{4}{*}{-} \\
      \hhline{~---~} & 4.26 & 45.17 & 6\,651$\pm$499 & \\
      \hhline{~---~} & 4.36 & 46.59 & 6\,420$\pm$420 & \\
      \hhline{~---~} & 4.42 & 46.69 &11\,083$\pm$615 & \\
      \hline
      \hline
    \end{tabular}
  \end{center}
\end{table*}
The fitted BFs are summarized in Table~\ref{table:simultaneous_fit},
together with the detection efficiencies and signal yields at each energy point.

\section{\boldmath CHARGED TRACK MULTIPLICITY OF
\texorpdfstring{$\eta_c$}{etac} INCLUSIVE DECAYS}
\label{sec:multiplicity}
The MC simulation for the inclusive $\eta_c$ decay has been
introduced in section~\ref{sec:detector_and_data_samples}.
The performance of the inclusive simulation, to some extent,
can be investigated by the consistency of the charged track multiplicity~\cite{R_scan, N_Jpsi, N_psip}.
Below, we introduce
how to obtain the true charged track multiplicity of $\eta_c$ inclusive decay.
An even number of charged tracks is generated
in an event due to the charge conservation,
while any number of charged tracks can be observed
due to the detector acceptance and reconstruction efficiency.
The observed charged track multiplicity of $\eta_c$ can be obtained
by fitting for the $\eta_c$ signal in the $\pi^+\pi^-\gamma$ recoil mass with the
number of extra candidate tracks required to be
0,~1,~2,~3,~$\cdots$, respectively.
To obtain the charged track multiplicity at the production level,
an unfolding method is employed based on an efficiency matrix,
whose matrix elements, $\epsilon_{ij}$,
represent the probabilities of an event generated with $j$ tracks
being observed with $i$ tracks.
The efficiency matrix is determined from the inclusive $\eta_c$ MC samples.
The unfolding of data is achieved by minimizing a $\chi^2$ value, defined as
\begin{equation}
\chi^2 =\sum\limits^{8}_{i=1}
  \frac{(N_i^{\rm obs}-\sum\limits_{j=0}^{8}\epsilon_{ij}\cdot N_j)^2}
  {(\sigma_i^{\rm obs})^2},
\label{eq:multiplicity_unfolding}
\end{equation}
where the values $N_i^{\rm obs}~(i=0,~1,~2,\cdots)$ are the observed multiplicities
of charged tracks in the data sample,
$\sigma_i^{\rm obs}$ are the corresponding uncertainties,
while $N_j~(j=0,~2,~4,\cdots)$ are the true
multiplicities of charged tracks at the production level in the data sample.
For simplicity, the events with eight or more tracks
are considered in a single value, $N_{\ge8}$,
so are the efficiencies, $\epsilon_{\ge8}$.

\begin{figure}[!htpb]
  \begin{center}
    \begin{overpic}[width=0.33\textwidth]
      {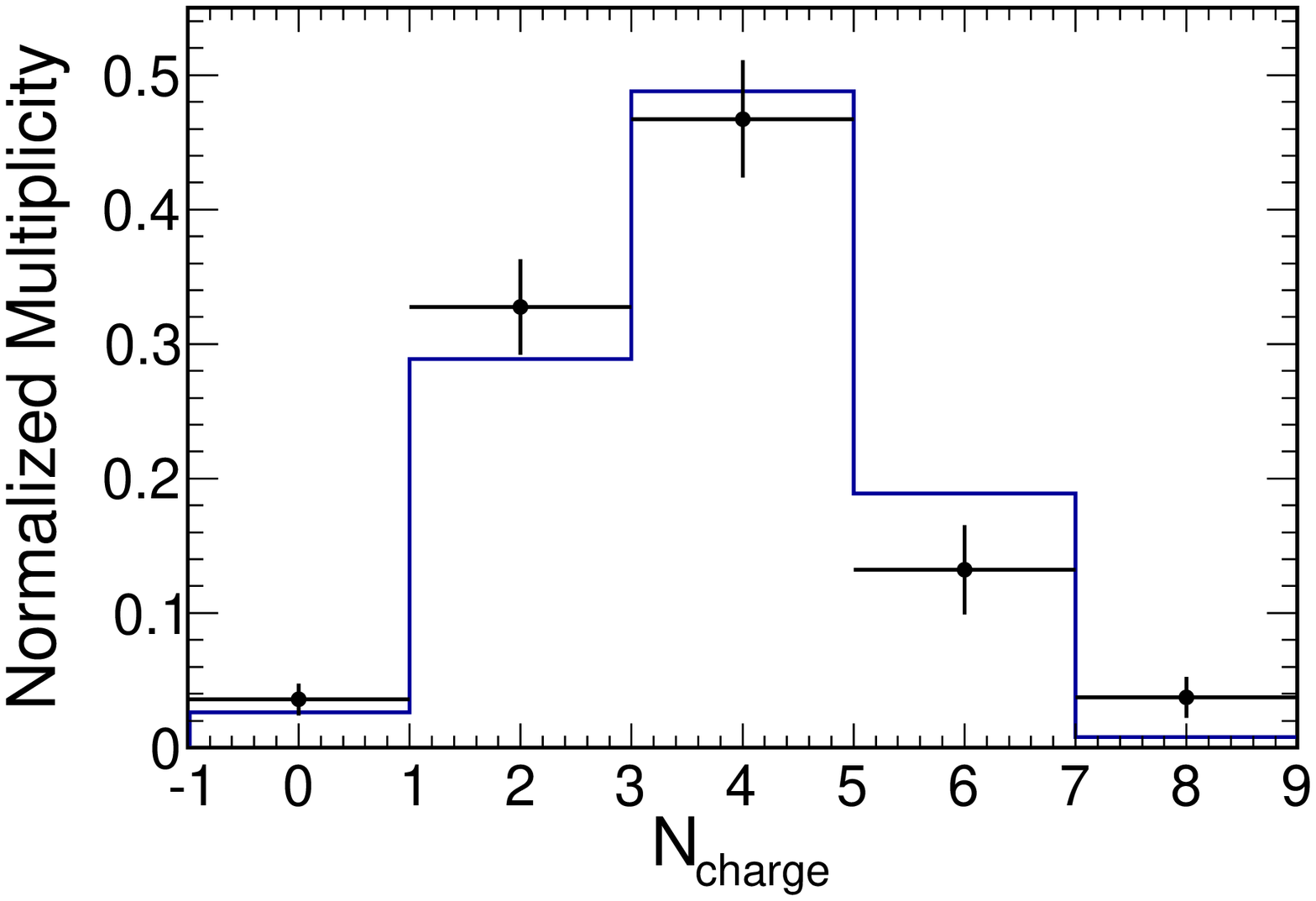}
    \end{overpic}
  \end{center}
  \caption{Normalized distributions of charged tracks multiplicities
  at the production level in $\eta_c$ decays,
  summed over all center-of-mass energies.
  The blue histogram represents results from MC,
  while black dots with error bar from data.
  The label 8 on the axis of $N_{\rm charge}$ means $N_{\rm charge}\ge8$.}
  \label{FIG:multiplicity}
\end{figure}
Figure~\ref{FIG:multiplicity} shows the charged track multiplicity distribution
of inclusive $\eta_c$ decays after combining the data at the four center-of-mass energies.
\begin{table*}[!hbp]\small
  \begin{center}
    \caption{The normalized multiplicity of $\eta_c$ at production level
    with systematic uncertainties.}
    \label{table:multiplicity}
    \begin{tabular}{c|c}
      \hline
      \hline
      $N_{\rm charge}$ & Normalized values\\
      \hline
      0      & $0.036\pm0.011\pm0.007$ \\
      \hline
      2      & $0.328\pm0.035\pm0.043$ \\
      \hline
      4      & $0.467\pm0.044\pm0.064$ \\
      \hline
      6      & $0.132\pm0.033\pm0.022$ \\
      \hline
      $\ge8$ & $0.037\pm0.015\pm0.009$ \\
      \hline
      \hline
    \end{tabular}
  \end{center}
\end{table*}
According to Eq.~\eqref{eq:multiplicity_unfolding},
the normalized numerical results are summarized in Table~\ref{table:multiplicity}.


\section{\boldmath SYSTEMATIC UNCERTAINTIES}
\subsection{\boldmath measurement of \texorpdfstring{${\rm BF}(\eta_c\to X)$}{etac->X}}
\label{sec:systematic_uncertainty_BF}
The systematic uncertainties on the BF measurements
for exclusive $\eta_c$ decays from different sources are described below
and listed in Table~\ref{table:sytematic_uncertainty}.
The total systematic uncertainty is determined by the sum in quadrature
of the individual values, assuming all sources to be independent.

\begin{table*}[!htbp]
  \begin{center}
    \caption{\small Relative systematic uncertainties~(in \%) in the branching fractions
    for the different final states of $\eta_c$ decays.}
    \label{table:sytematic_uncertainty}
    \begin{tabular}{c|c|c|c|c|c}
      \hline
      \hline
      \multicolumn{2}{c|}{Category~(\%)} &
      $\eta_c\to K^+K^-\pi^0$ & $\eta_c\to K^0_S K^{\pm} \pi^{\mp}$ &
      $\eta_c\to 2(\pi^+\pi^-\pi^0)$ & $\eta_c \to p \bar{p}$ \\
      \hhline{------}
      \multicolumn{2}{c|}{Tracking} & 4.0 & 3.0 & 4.0 & 4.0 \\
      \hhline{------}
      \multicolumn{2}{c|}{PID} & 2.0 & 2.0 & 4.0 & 2.0 \\      \hline
      \multicolumn{2}{c|}{$\pi^0$ reconstruction} & 3.75 & - & 3.23 & - \\
      \hline
      \multicolumn{2}{c|}{Kinematic Fit} & 0.46 & 0.30 & 1.09 & 0.07 \\
      \hline
      \multicolumn{2}{c|}{$K^0_S$ reconstruction} & - & 1.2 & - & - \\
      \hline
      \multicolumn{2}{c|}{MC model} & 0.85 & 0.79 & 1.49 & 0.73  \\
      \hline
      \multicolumn{2}{c|}{$h_c$ mass window} & 1.93 & 2.35 & 3.01 & 5.91 \\
      \hline
      \multirow{8}{*}{Fitting} &
      Fitting range & 5.62 & 5.21 & 6.56 & 3.65  \\
      \hhline{~-----} & Background shape~(exclusive) & 0.60 & 0.63 & 5.12 & 8.37  \\
      \hhline{~-----} & Sidebands range~(inclusive) & 1.17 & 1.26 & 1.25 & 1.14  \\
      \hhline{~-----} & Background form~(inclusive) & 2.63 & 2.73 & 2.67 & 2.71  \\
      \hhline{~-----} & Mass resolution & 0.06 & 0.10 & 0.14 & 0.10  \\
      \hhline{~-----} & Resonant parameters of $\eta_c$ & 0.81 & 0.81 & 0.38 & 0.79 \\
      \hhline{~-----} & Damping factors & 0.89 & 1.57 & 1.09 & 1.74 \\
      \hline
      \multicolumn{2}{c|}{Total} & 9.0 & 7.7 & 11.6 & 12.3 \\
      \hline
      \hline
    \end{tabular}
  \end{center}
\end{table*}

\subsubsection{MDC tracking and PID}
\label{sec:systematic_uncertainty_BF:MDC_PID}
The uncertainty from the tracking efficiency and PID
for the two soft pions in the process $e^+e^-\to\pi^+\pi^-h_c$ cancels
since the BFs are measured by a relative method,
as mentioned in the introduction.
We only consider the uncertainty from tracking efficiency and PID
of the $\eta_c$ decay products.
The involved charged tracks are pions~(not including the pions from $K_S^0$ decay), kaons, and protons.
Their uncertainties are studied with different control samples,
$e^+e^-\to\pi^+\pi^-K^+K^-$ for pions and kaons,
$e^+e^-\to p\pi^-\bar{p}\pi^+$($e^+e^-\to p\pi^-\bar{p}\pi^+\pi^+\pi^-$) for protons,
The uncertainties from tracking efficiency are 1\% for each pion,
and 2\% for each kaon or proton.
The uncertainties for PID are 1\% for each pion, kaon or proton.

\subsubsection{\texorpdfstring{$\pi^0$}{pi0} reconstruction}
The systematic uncertainty from $\pi^0$ reconstruction is studied with
$\psi(3686)\to \pi^0\pi^0 J/\psi$ using $1.06\times10^8$~$\psi(3686)$ events
and $e^+e^-\to \omega\pi^0\to\pi^+\pi^-\pi^0\pi^0$
using a data sample of $2.93~\rm{fb}^{-1}$ collected at the $\psi(3770)$ resonance.
The uncertainty as a function of $\pi^0$ momentum is determined.
The uncertainty from $\pi^0$ reconstruction is calculated with the function,
according to the momentum distribution of the $\pi^0$ in the decays studied.

\subsubsection{Kinematic fit}
The systematic uncertainty from the kinematic fit is estimated
by correcting the helix parameters of the charged tracks in the MC simulation~\cite{kinematic_fit}.
The differences in the detection efficiency between the MC samples with and without the corrections
are taken as the uncertainties due to the kinematic fit.

\subsubsection{\texorpdfstring{$K_S^0$}{Ks} reconstruction}
The $K^0_S$ reconstruction is studied with two control samples,
$J/\psi\to K^{*\pm} \bar{K}^{\mp}$ and $J/\psi\to\phi K^0_S K^{\pm}\pi^{\mp}$.
The difference in the $K^0_S$ reconstruction
efficiency between the MC simulation and the data is 1.2\%~\cite{ks},
which is taken as the uncertainty due to $K_S^0$ reconstruction.

\subsubsection{MC model}
\label{sec:systematic_uncertainty_BF:MC_model}
In the MC simulation, the process $e^+e^-\to\pi^+\pi^- h_c$ is modeled with a phase space~({\sc
  phsp}) distribution.
In fact, there is a confirmed intermediate state $Z_c(4020)$ and a potential intermediate state $Z_c(3900)$,
in the $\pi^+\pi^- h_c$ final state. The uncertainty caused by the intermediate states is estimated
by mixing the MC events including $Z_c(4020)$/$Z_c(3900)$ component according to the measured fractions~\cite{guo_yuping_zc,zc_3900}.
The difference in the detection efficiency is taken as the uncertainty.

For the exclusive $\eta_c$ decay modes, intermediate resonant states may affect the detection efficiency.
MC samples related to $\eta_c$ multi-body decays are generated by sampling according to the invariant mass distribution
or mixing the known intermediate states, or changing the decay model used in the MC simulation.
The difference in the efficiency with and without intermediate states is taken as the uncertainty.

The uncertainty due to the inconsistency between data and MC simulation
on the charged track multiplicity in inclusive $\eta_c$ decays
is estimated based on the multiplicity obtained by the unfolding method mentioned in Sec.~\ref{sec:multiplicity}.
The detection efficiency for inclusive decay can also be re-calculated with the following formula,
  $$\epsilon_{\rm inclusive} = \sum\limits_{j}(N_j\sum\limits_{i}\epsilon_{ij}),$$
where $N_i$ are the normalized multiplicities in data,
listed in Table~\ref{table:multiplicity}, and 
$\epsilon_{ij}$ are the elements of the efficiency matrix in Eq.~\eqref{eq:multiplicity_unfolding}.
The differences between this result and the original one are taken into account in the simultaneous fit.
It is found that the influence on ${\rm BF}(\eta_c\to X)$ is negligible.

\subsubsection{\texorpdfstring{$h_c$}{hc} mass window}
\label{sec:systematic_uncertainty_BF:hc_window}
The uncertainty from the $h_c$ mass window is estimated by randomly changing the low and high boundaries of the $h_c$ signal region
in the ranges of $[3.512,~3.518]~\rm{GeV}/c^2$ and $[3.532,~3.538]~\rm{GeV}/c^2$
and fitting the spectrum with efficiencies estimated in the corresponding intervals.
The procedure is repeated for 800 times, and the distributions of the fitted BFs follow Gaussian functions.
The obtained standard deviations are taken as the uncertainties
due to the $h_c$ mass window selection.

\subsubsection{Fit procedure}
\label{sec:systematic_uncertainty_BF:fitting}
This uncertainty arises from the fit range, the background shape,
the mass resolution, the parameters of the $\eta_c$ resonance,
the efficiency curves, and the damping factor.

The uncertainty from the fit range is estimated
by randomly changing the lower side in the range of $[2.540,~2.555]$~GeV/$c^2$
and higher side in $[3.200,~3.215]$~GeV/$c^2$
and repeating the fit for 800 times.
The root mean square~(RMS) of the resulting distributions are taken
as the systematic uncertainties from the fit range.

The uncertainty due to the assumed background shape in the exclusive modes is estimated by changing
the order of the Chebychev polynomial functions.
For the inclusive decay mode, the $h_c$ sidebands need to be considered as well,
whose systematic uncertainty  is estimated by randomly changing the left and right margins
of the lower and upper sidebands and repeating the fit.
The procedure is performed 800 times.
The left and right margins of the sidebands are changed
in the ranges of $[3.496,~3.450],~[3.503,~3.507]~\rm{GeV}/c^2$
and $[3.543,~3.547],~[3.548,~3.552]~\rm{GeV}/c^2$ for the lower and upper sideband
regions, respectively. The distributions of the fitted results follow Gaussian functions, and
the standard deviations are taken as the uncertainties from the $h_c$ sidebands selection.
The uncertainty from the  polynomial is estimated by changing the order of the polynomial.

The discrepancy between data and MC simulation on detection resolution is estimated by a control sample,
$\psi(2S)\to \pi^+\pi^- J/\psi$, $J/\psi\to\gamma\eta^\prime$, $\eta^\prime\to \gamma\pi^+\pi^-$.
By fitting the $\eta^\prime$ signals, we can obtain the mass resolution for both data and MC.
We change the mass resolutions
according to the result obtained from control sample
to re-fit the $RM(\gamma\pi^+\pi^-)$.
The differences on the BFs with and without changing the mass resolution are taken as the systematic uncertainties.

The $\eta_c$ resonance parameters are fixed to the world average values in the fit.
We change these values by $\pm 1\sigma$, and the larger difference is taken as the uncertainty.

The efficiency curves, as shown in Fig.~\ref{FIG:efficiency_curves},
change slowly with $RM(\pi^+\pi^-\gamma)$.
We find only a very small change in results when constant efficiencies are used.
Therefore, the uncertainties due to efficiencies can be neglected.

The uncertainty from the damping factor is estimated by using an alternative form of
the damping factor, which is used in the CLEO's published paper~\cite{cleo}.
The differences between the results with the two forms of damping factor are taken as the systematic uncertainty.

\subsection{Charged track multiplicity}
The systematic uncertainties on the charged track multiplicity
in $\eta_c$ inclusive decay from different sources are described below
and listed in Table~\ref{table:sytematic_uncertainty_multiplicity}.
They are estimated in a similar way as introduced in Sec.~\ref{sec:systematic_uncertainty_BF}.
The total systematic uncertainty is determined
by the sum in quadrature of the individual values,
assuming that all the sources are independent.

\begin{table*}[!htbp]
  \begin{center}
    \caption{\small Systematic uncertainties~(\%) in the multiplicity of $\eta_c$.}
    \label{table:sytematic_uncertainty_multiplicity}
    \begin{tabular}{c|c|c|c|c|c|c}
      \hline
      \hline
      \multicolumn{2}{c|}{Category~(\%)} &
      $N_0$ & $N_2$ & $N_4$ & $N_6$ & $N_{\ge 8}$\\
      \hhline{-------}
      \multicolumn{2}{c|}{Tracking} & 2.0 & 2.0 & 2.0 & 2.0 & 2.0\\
      \hhline{-------}
      \multicolumn{2}{c|}{PID} & 2.0 & 2.0 & 2.0 & 2.0 & 2.0\\
      \hhline{-------}
      \multirow{2}{*}{MC model} & intermediate states & 4.19 & 3.46 & 5.22 & 7.47 & 7.47 \\
      \hhline{~------} & multiplicity & 10.40 & 10.60 & 11.76 & 9.31 & 8.87\\
      \hline
      \multicolumn{2}{c|}{$h_c$ mass window} & 11.70 & 3.54 & 3.01 & 5.91 & 15.26 \\
      \hline
      \multirow{5}{*}{Fit} &
      Fitting range & 5.92 & 3.84 & 1.13 & 4.28 & 6.34\\
      \hhline{~------} & Background shape& 8.04 & 3.41 & 1.96 & 8.96 & 11.80\\
      \hhline{~------} & Mass resolution & 0.14 & 0.10 & 0.01 & 0.32 & 0.46\\
      \hhline{~------} & Resonant parameters of $\eta_c$ & 0.68 & 0.34 & 0.44 & 0.65& 0.85\\
      \hhline{~------} & Damping factors & 1.35 & 0.34 & 0.34 & 0.56& 4.10\\
      \hline
      \multicolumn{2}{c|}{Total} & 19.3 & 13.1 & 13.7 & 16.9 & 23.9\\
      \hline
      \hline
    \end{tabular}
  \end{center}
\end{table*}
\subsubsection{MDC tracking and PID}
The uncertainties from MDC tracking and PID are the same as those in the measurement of ${\rm
  BF}(\eta_c\to X)$.
\subsubsection{\texorpdfstring{$h_c$}{hc} mass window}
The uncertainties are estimated by changing the $h_c$ mass window from $[3.515, 3.535]~{\rm GeV}/c^2$
to $[3.518, 3.532]~{\rm GeV}/c^2$ and $[3.512, 3.538]~{\rm GeV}/c^2$.
The largest changes on the multiplicity are taken as the uncertainty.
\subsubsection{MC model}
Similar to that in measurement of ${\rm BF}(\eta_c\to X)$,
the uncertainty due to MC model mainly comes from the potential intermediate states and the inclusive $\eta_c$ decay.
The uncertainty from the former is estimated as before,
while the latter is estimated by removing the unknown modes simulated by {\sc lundcharm} model,
and only considering the known $\eta_c$ decay modes.  

\subsubsection{Fit}
The uncertainties due to the fit to the recoil mass spectra of $\pi^+\pi^-\gamma$
are evaluated by varying the fit range, sideband ranges, mass resolution,
resonant parameters of $\eta_c$, and damping factors used in the fit,
in similar ways as introduced in Sec.~\ref{sec:systematic_uncertainty_BF}.
The spreads of the results obtained with the alternative assumptions are used
to assign the systematic uncertainties.
\section{\boldmath SUMMARY}

\begin{table*}[!htbp]
  \begin{center}
    \caption{\small Measured BFs of
    $\eta_c\to K^+K^-\pi^0$, $K^0_S K^{\pm}\pi^{\mp}$,
    $2(\pi^+\pi^-\pi^0)$, and $p \bar{p}$
    with statistical~(the first ones)
    and systematic~(the second ones) uncertainties.
    The third uncertainties in the results from Ref.~\cite{guo_aiqiang_etac}
    are the systematic uncertainties due to the uncertainty of
    ${\rm BF}(\psi(3686)\to\pi^0 h_c)\times {\rm BF}(h_c\to\gamma\eta_c)$.
    The combined results from PDG are listed in the last column,
    among which ${\rm BF}(\eta_c\to K\bar{K}\pi)$ is provided.}
    \label{table:results_summary}
    \begin{tabular}{c|c|c|c}
      \hline
      \hline
      Final states & BF~(\%) & BF~(\%) from Ref.~\cite{guo_aiqiang_etac} & BF~(\%) from PDG~\cite{pdg}  \\
      \hline
      $K^+ K^- \pi^0$         & $1.15 \pm0.12 \pm0.10$      & $1.04\pm0.17\pm0.11\pm0.10$   & \multirow{2}{*}{$7.3\pm0.5~(K\bar{K}\pi)$}\\
      $K_S^0K^{\pm}\pi^{\mp}$ & $2.60 \pm0.21 \pm0.20$      & $2.60\pm0.29\pm0.34\pm0.25$   & \\
      $2(\pi^+\pi^-\pi^0)$    & $15.3 \pm1.8  \pm1.8$ &      $17.23\pm1.70\pm2.29\pm1.66$    & $17.4\pm3.3$\\
      $p\bar{p}$              & $0.120\pm0.026\pm0.015$   & $0.15\pm0.04\pm0.02\pm0.01$   & $0.152\pm0.016$\\
      \hline
      \hline
    \end{tabular}
  \end{center}
\end{table*}

In summary, with the data samples collected at $\sqrt{s} = $4.23, 4.26, 4.36, and 4.42~GeV,
by comparing the exclusive and inclusive decays of $\eta_c$,
we determine the BFs for
$\eta_c\to K^+K^-\pi^0$, $K^0_S K^{\pm}\pi^{\mp}$,
$2(\pi^+\pi^-\pi^0)$, and $p \bar{p}$
via $e^+e^-\to\pi^+\pi^-h_c$, $h_c\to\gamma\eta_c$.
The results are presented in Table~\ref{table:results_summary};
they agree with previous measurements
by BESIII~\cite{guo_aiqiang_etac} within uncertainties,
while the accuracy of these BFs is improved.
With this improved accuracy,
the measurements of the {\it M}1 transitions of
$J/\psi\to \gamma\eta_c$ and $\psi(3686)\to\gamma\eta_c$ can be more precise,
since such measurements provide combined results of
${\rm BF}(J/\psi(\psi(3686))\to\gamma\eta_c)\times{\rm BF}(\eta_c\to X)$.

Moreover, the charged track multiplicity of $\eta_c$ inclusive 
decay at production level is quantitatively presented for the first time
in Table~\ref{table:multiplicity}. 
The good consistency between data and MC simulation for this charged track multiplicity
indicates that the current MC simulation works generally well.
With this charged track multiplicity, many
studies with  $\eta_c$ in the final state~\cite{b2} are possible with higher precision than
previously.
\section{ACKNOWLEDGEMENT}
The BESIII collaboration thanks the staff of BEPCII and the IHEP computing center for their strong support. This work is supported in part by National Key Basic Research Program of China under Contract No. 2015CB856700; National Natural Science Foundation of China (NSFC) under Contracts Nos. 11335008, 11425524, 11625523, 11635010, 11735014; the Chinese Academy of Sciences (CAS) Large-Scale Scientific Facility Program; the CAS Center for Excellence in Particle Physics (CCEPP); Joint Large-Scale Scientific Facility Funds of the NSFC and CAS under Contracts Nos. U1532257, U1532258, U1732263; CAS Key Research Program of Frontier Sciences under Contracts Nos. QYZDJ-SSW-SLH003, QYZDJ-SSW-SLH040; 100 Talents Program of CAS; INPAC and Shanghai Key Laboratory for Particle Physics and Cosmology; German Research Foundation DFG under Contract Nos. Collaborative Research Center CRC 1044, FOR 2359; Istituto Nazionale di Fisica Nucleare, Italy; Koninklijke Nederlandse Akademie van Wetenschappen (KNAW) under Contract No. 530-4CDP03; Ministry of Development of Turkey under Contract No. DPT2006K-120470; National Science and Technology fund; The Swedish Research Council; U. S. Department of Energy under Contracts Nos. DE-FG02-05ER41374, DE-SC-0010118, DE-SC-0010504, DE-SC-0012069; University of Groningen (RuG) and the Helmholtzzentrum fuer Schwerionenforschung GmbH (GSI), Darmstadt.

\end{document}